\documentclass[twocolumn,superscriptaddress,showpacs,preprintnumbers,amsmath,amssymb]{revtex4}
\usepackage{graphicx,color}% Include figure files
\usepackage{dcolumn}% Align table columns on decimal point
\usepackage{bm}% bold math
\usepackage{CJK}
\usepackage{mathrsfs}

\newcommand{\lb}{\left(}
\newcommand{\rb}{\right)}
\newcommand{\ls}{\left[}
\newcommand{\rs}{\right]}

\newcommand{\ff}[1]{\frac{1}{#1}}

\begin{document}
\begin{CJK*}{GBK}{song}

\title{Pseudospin symmetry in supersymmetric quantum mechanics:
    II. Spin-orbit effects}% Force line breaks with \\

\author{Shihang Shen}
 \affiliation{State Key Laboratory of Nuclear Physics and Technology, School of Physics,
Peking University, Beijing 100871, China}

\author{Haozhao Liang}
 \affiliation{RIKEN Nishina Center, Wako 351-0198, Japan}
 \affiliation{State Key Laboratory of Nuclear Physics and Technology, School of Physics,
Peking University, Beijing 100871, China}

\author{Pengwei Zhao}
 \affiliation{State Key Laboratory of Nuclear Physics and Technology, School of Physics,
Peking University, Beijing 100871, China}

\author{Shuangquan Zhang}
 \affiliation{State Key Laboratory of Nuclear Physics and Technology, School of Physics,
Peking University, Beijing 100871, China}

\author{Jie Meng}
 \affiliation{State Key Laboratory of Nuclear Physics and Technology, School of Physics,
Peking University, Beijing 100871, China}
 \affiliation{School of Physics and Nuclear Energy Engineering, Beihang University,
              Beijing 100191, China}
 \affiliation{Department of Physics, University of Stellenbosch, Stellenbosch, South Africa}

\date{\today}

\begin{abstract}
Following a previous paper [Haozhao Liang \textit{et al.}, Phys. Rev. C \textbf{87}, 014334 (2013)], we discuss the spin-orbit effects on the pseudospin symmetry (PSS) within the framework of supersymmetric quantum mechanics.
By using the perturbation theory, we demonstrate that the perturbative nature of PSS maintains when a substantial spin-orbit potential is included. With the explicit PSS-breaking potential, the spin-orbit effects on the pseudospin-orbit splittings are investigated in a quantitative way.
\end{abstract}

\maketitle

%==================Introduction======================================
\section{Introduction}

In a previous paper~\cite{Liang2013}, we reported an investigation on the origin of pseudospin symmetry (PSS) and its breaking mechanism by combining the similarity renormalization group (SRG) technique, supersymmetry (SUSY) quantum mechanics, and perturbation theory. Under the lowest-order approximation in transforming a Dirac equation into a diagonal form by means of the SRG method, the Schr\"odinger equation without a spin-orbit (SO) term was investigated in an explicit and quantitative way~\cite{Liang2013}.
Since the remarkable spin-orbit splitting is one of the most important features in nuclear physics~\cite{Haxel1949,Goeppert-Mayer1949} and the SO term represents a relativistic correction to the Schr\"odinger equation for spin $1/2$ particles~\cite{Bjorken1964}, it is important to quantitatively investigate the spin-orbit effects on the PSS breaking with the framework of SRG, SUSY and perturbation theory.
This becomes the main task of the present paper.

The concept of pseudospin symmetry~\cite{Arima1969,Hecht1969} was introduced to explain the near degeneracy between two single-nucleon states with the quantum numbers ($n-1, l + 2, j = l + 3/2$) and ($n, l, j=l + 1/2$).
They are regarded as the pseudospin doublets as ($\tilde{n}=n-1,\tilde{l}=l+1,j=\tilde{l}\pm1/2$).
Since the suggestion of PSS in atomic nuclei, there have been comprehensive efforts to understand its origin.
In 1997, the PSS was shown to be a symmetry of the Dirac Hamiltonian, and the equality in magnitude but difference in sign of the scalar potential $\mathcal{S}(\mathbf{r})$ and vector potential $\mathcal{V}(\mathbf{r})$ was suggested as the exact PSS limit~\cite{Ginocchio1997}.
A more general condition is $d(\mathcal{S}+\mathcal{V})/dr=0$~\cite{Meng1998,Sugawara-Tanabe1998}, which can be better satisfied in exotic nuclei with diffuse potentials~\cite{Meng1999}.
On the other hand, since there exist no bound nuclei within such PSS limit, the non-perturbative or dynamical nature of PSS in realistic nuclei was suggested~\cite{Alberto2001}. An extensive review on the pseudospin symmetry investigation has been given in the precedent paper~\cite{Liang2013}, and the readers are referred to Refs.~\cite{Lu2012,Castro2012,Chen2012,Jolos2012,Alberto2013,Lu2013Arxiv} for recent progresses.

Recently, the perturbation theory was used to investigate the symmetries of the Dirac Hamiltonian and their breaking in realistic nuclei~\cite{Liang2011,Li2011}, which provides a clear and quantitative way for investigating the perturbative nature of PSS.
On the other hand, the SUSY quantum mechanics can provide a PSS-breaking potential without singularity~\cite{Typel2008}, and naturally interpret the unique feature that all states with $\tilde{l}>0$ have their own pseudospin partners except for the intruder states~\cite{Leviatan2004,Typel2008,Leviatan2009}.
Furthermore, the SRG technique fills the gap between the perturbation calculations and the SUSY descriptions by transforming the Dirac Hamiltonian into a diagonal form which keeps every operator Hermitian~\cite{Guo2012,Li2013}.
Therefore, we deem it promising to understand the PSS and its breaking mechanism in a fully quantitative way by combining the SRG technique, SUSY quantum mechanics, and perturbation theory.
This initiated the investigations in Ref.~\cite{Liang2013}.

As an indispensable step forward, one should take the SO term into account because it plays a crucial role in nuclear shell structure. In this paper, the SO term in the SUSY representation will be discussed and its effects on the PSS breaking will be investigated quantitatively.

%==================Theoretical Framework=============================
\section{Theoretical Framework}\label{Sect:II}

In this Section, we will mainly focus on SUSY quantum mechanics for the spin-orbit terms in the Schr\"odinger equations obtained by SRG. For the formalism concerning the related SRG technique, one is referred to Refs.~\cite{Guo2012,Liang2013} for details.

Within the relativistic scheme, the Dirac Hamiltonian for nucleons reads
\begin{equation}\label{Eq:Dirac}
H_D =   \bm{\alpha\cdot p} + \beta (M+\mathcal{S}) + \mathcal{V}
\end{equation}
where $\bm{\alpha}$ and $\beta$ are the Dirac matrices, $M$ is the mass of nucleon, $\mathcal{S}$ and $\mathcal{V}$ are the scalar and vector potentials respectively.
Using the SRG technique, the eigenequations for nucleons in the Fermi sea include the SO term $\frac{\kappa\Delta'}{r}\lb-\frac{1}{4M^2}+\frac{\mathcal{S}}{2M^3}+\cdots\rb$, which is proportional to $\frac{\kappa\Delta'}{r}$ and similar to the effective SO term $\frac{\kappa\Delta'}{r}\frac{1}{(E+2M-\Delta)^2}$ obtained by reducing the Dirac equation to a Schr\"odinger-like equation for the upper component~\cite{Guo2012,Liang2013}. Here $\Delta'(\bm{r})$  stands for the derivative of $\Delta$ with respect to $r$ and $E$ is the single-particle energy excluding the mass of nucleon.
Focusing on the SUSY representation for the SO terms proportional to $\frac{\kappa\Delta'}{r}$, it is sufficient to take the lowest ($1/M^2$)th order as an example. The corresponding Schr\"odinger equations read
\begin{equation}\label{Eq:Schrodinger}
\left[-\frac{1}{2M}\nabla^2 + V(\bm{r}) - \frac{\kappa}{r}\frac{\Delta'(\bm{r})}{4M^2} \right] \psi(\bm{r}) = E \psi(\bm{r})
\end{equation}
with $V = \mathcal{V} + \mathcal{S}, \Delta = \mathcal{V} - \mathcal{S}$.
By assuming the spherical symmetry, the radial Schr\"odinger equations are expressed as
\begin{equation}\label{Eq:Schr}
    H R(r) = E R(r)
\end{equation}
with the single-particle Hamiltonian
\begin{equation}\label{Eq:Hka}
    H = -\frac{1}{2M}\frac{d^2}{dr^2} + \frac{\kappa(\kappa+1)}{2Mr^2} + V(r) - \frac{\kappa}{r}\frac{\Delta'}{4M^2}.
\end{equation}
The good quantum number $\kappa$ is defined as $\kappa=\mp(j+1/2)$ for $j=l\pm1/2$.

For applying the SUSY quantum mechanics to the Schr\"{o}dinger equations shown in Eq.~(\ref{Eq:Schr}), we rewrite the Hamiltonian in Eq.~(\ref{Eq:Hka}) as
\begin{equation}\label{Eq:HkaNew}
    H = -\frac{1}{2M}\frac{d^2}{dr^2} + \frac{\kappa(\kappa+1)}{2Mr^2} + V(r) + \frac{\kappa}{Mr}U(r),
\end{equation}
where $U(r)=-\Delta'/(4M)$.

In the SUSY framework, a couple of Hermitian conjugate first-order differential operators are defined as
\begin{equation}
B_\kappa^+ = \ls Q_\kappa(r)-\frac{d}{dr}\rs \frac{1}{\sqrt{2M}},\quad B_\kappa^- = \frac{1}{\sqrt{2M}}\ls Q_\kappa(r)+\frac{d}{dr}\rs,
\end{equation}
where the $Q_\kappa(r)$ are the so-called superpotentials to be determined~\cite{Cooper1995,Typel2008}.
In order to explicitly identify the $\kappa(\kappa+1)$ structure and the SO term shown in Eq.~(\ref{Eq:HkaNew}), one can  introduce the reduced superpotentials $q_\kappa(r)$ as
\begin{equation}
    q_\kappa(r) = Q_\kappa(r) - \frac{\kappa}{r} - U(r).
\end{equation}
In such a way, the SUSY partner Hamiltonians $H_1$ and $H_2$ can be expressed as
\begin{widetext}
\begin{subequations}\label{Eq:BB}
\begin{eqnarray}
    H_1(\kappa) &=& B^+_\kappa B^-_\kappa = \ff{2M}\ls-\frac{d^2}{dr^2}+\frac{\kappa(\kappa+1)}{r^2}+q_\kappa^2+2q_\kappa U+U^2+\frac{2\kappa}{r}q_\kappa-q'_\kappa-U'\rs+\frac{\kappa}{Mr}U,\notag \\ \label{Eq:BB1}\\
    H_2(\kappa) &=& B^-_\kappa B^+_\kappa = \ff{2M}\ls-\frac{d^2}{dr^2}+\frac{\kappa(\kappa-1)}{r^2}+q_\kappa^2+2q_\kappa U+U^2+\frac{2\kappa}{r}q_\kappa+q'_\kappa+U'\rs+\frac{\kappa}{Mr}U.\notag \\ \label{Eq:BB2}
\end{eqnarray}
\end{subequations}
\end{widetext}
It can be seen that not only does the $\kappa(\kappa+1)$ structure appear in $H_1$ but also the $\kappa(\kappa-1)$ structure explicitly appears in the SUSY partner Hamiltonian $H_2$.
The so-called pseudospin centrifugal barrier (PCB) terms $\kappa(\kappa-1)/(2Mr^2)$ are identical for the pseudospin doublets $a$ and $b$ with $\kappa_a + \kappa_b = 1$.

By combining Eq.~(\ref{Eq:HkaNew}) and Eq.~(\ref{Eq:BB1}), one obtains the first-order differential equations for the reduced superpotentials $q_\kappa(r)$,
\begin{equation}\label{Eq:qr}
\frac{1}{2M}\ls (q_\kappa+U)^2+\frac{2\kappa}{r}q_\kappa-(q_\kappa+U)'\rs +e(\kappa) = V(r),
\end{equation}
where the constants $e(\kappa)$ are the so-called energy shifts \cite{Cooper1995,Typel2008,Liang2013}.
The $\kappa$-dependent energy shifts are determined in the same way as that in Ref.~\cite{Liang2013}:
(1) For the case of $\kappa<0$,
\begin{equation}
    e(\kappa_a) = E_{1\kappa_a},
\end{equation}
since the SUSY is exact.
(2) For the case of $\kappa>0$,
\begin{equation}
    e(\kappa_a) = \ls 2V-\frac{U^2-U'}{M}\rs_{r=0} - e(\kappa_b),
\end{equation}
with $\kappa_a+\kappa_b=1$ for pseudospin doublets, which makes the pseudospin-orbit (PSO) potentials vanish as $r\rightarrow0$.
Such vanishing behavior is similar to that of the usual surface-peak-type SO potentials, where $\lim_{r\rightarrow0}U(r) = \lim_{r\rightarrow0}U'(r) = 0$.

In addition, the asymptotic behaviors of the reduced superpotentials $q_\kappa(r)$ in Eq.~(\ref{Eq:qr}) are as follows:
At large radius, for potentials $\lim_{r\rightarrow\infty}V(r)=\lim_{r\rightarrow\infty}U(r)=0$, $q_\kappa(r)$ becomes a constant as
\begin{equation}\label{Eq:qinf}
    \lim_{r\rightarrow\infty}q_\kappa(r) = \sqrt{-2Me(\kappa)};
\end{equation}
At small radius, for any regular potentials $V(r)$ and $U(r)$, it requires $q_\kappa(0)=0$, and also
\begin{equation}\label{Eq:q0}
    \lim_{r\rightarrow0}q_{\kappa}(r)=\frac{2M(e(\kappa)-V)}{(1-2\kappa)}r
\end{equation}
as a linear function of $r$.

As shown in Eq.~(\ref{Eq:BB2}), the SUSY partner Hamiltonian reads
\begin{align}\label{Eq:Htil}
    \tilde{H}(\kappa)&=H_2(\kappa)+e(\kappa)\nonumber\\
    &= -\frac{d^2}{2Mdr^2}+\frac{\kappa(\kappa-1)}{2Mr^2}+ \tilde{V}_\kappa(r)+ \frac{\kappa}{Mr}U(r),
\end{align}
with $\tilde{V}_\kappa(r) = V(r) + [q_\kappa'(r)+U'(r)]/M$.
For the perturbation analysis, the Hamiltonian $\tilde{H}$ is further expressed as~\cite{Liang2013}
\begin{equation}\label{Eq:HH0}
  \tilde{H} = \tilde{H}^{\rm PSS}_0 + \tilde{W}^{\rm PSS},
\end{equation}
where $\tilde{H}^{\rm PSS}_0$ and $\tilde{W}^{\rm PSS}$ are the corresponding PSS-conserving and PSS-breaking terms, respectively. By requiring that $\tilde{W}^{\rm PSS}$ should be proportional to $\kappa$~\cite{Liang2013}, which is similar to the case of the
SO term in the normal scheme, one has
\begin{subequations}
\begin{eqnarray}
    \tilde{H}^{\rm PSS}_0&=&\ff{2M}\ls-\frac{d^2}{dr^2}+\frac{\kappa(\kappa-1)}{r^2}\rs + \tilde{V}_{\rm PSS}(r),\\
    \tilde{W}^{\rm PSS}&=&\kappa \tilde{V}_{\rm PSO}(r).
\end{eqnarray}
\end{subequations}
Finally, for a given pair of pseudospin doublets $a$ and $b$ with $\kappa_a+\kappa_b=1$, the PSO potential $\tilde{V}_{\rm PSO}(r)$ can be uniquely determined as
\begin{equation}\label{Eq:VPSO}
  \tilde{V}_{\rm PSO}(r)
    =\frac{\tilde{H}(\kappa_a)-\tilde{H}(\kappa_b)}{\kappa_a-\kappa_b}
    =\frac{1}{M}\frac{q'_{\kappa_a}(r) - q'_{\kappa_b}(r)}{\kappa_a-\kappa_b}
        + \frac{1}{Mr}U(r).
\end{equation}

In this paper, we use a tilde to denote the operators, potentials, and wave functions belonging to the representation of $\tilde{H}$.

%==================Results and discussion===============================================
\section{Results and discussion}\label{Sect:III}

In the following calculations, the mass of nucleon is taken as $M=939.0$~MeV, the central potential $V(r)$ in Eq.~(\ref{Eq:HkaNew}) is chosen as a Woods-Saxon form
\begin{equation}
    V(r) = \frac{V_0}{1+e^{(r-R)/a}},
\end{equation}
and the corresponding $\Delta(r)$ in the SO term reads
\begin{equation}
    \Delta(r) = \frac{-\lambda V_0}{1+e^{(r-R^{\rm ls})/a^{\rm ls}}},
\end{equation}
with the parameters $V_0=-63.297$~MeV, $\lambda=11.12$, $R=6.278$~fm, $R^{\rm ls}=5.825$~fm, $a=0.615$~fm, and $a^{\rm ls}=0.648$~fm, which are taken from Ref.~\cite{Koepf1991} for the neutron potential of $^{132}$Sn.
The radial Schr\"odinger equations are solved in coordinate space by the shooting method~\cite{Meng1998NPA} within a spherical box with radius $R_{\rm box} = 20$~fm and mesh size $dr = 0.05$~fm.
Comparing with the self-consistent calculations with relativistic mean-field theory~\cite{Liang2011b}, the higher-order SO terms and the effective mass terms are neglected in the present discussions. However, the SO splittings for $3p$ and $2f$ spin doublets thus obtained are $0.613$ and $1.723$~MeV, respectively, while the corresponding experiment data are $0.509$ and $2.005$~MeV~\cite{Jones2010}. This indicates the SO effects discussed here are close to reality.

\begin{figure}
\includegraphics[width=8cm]{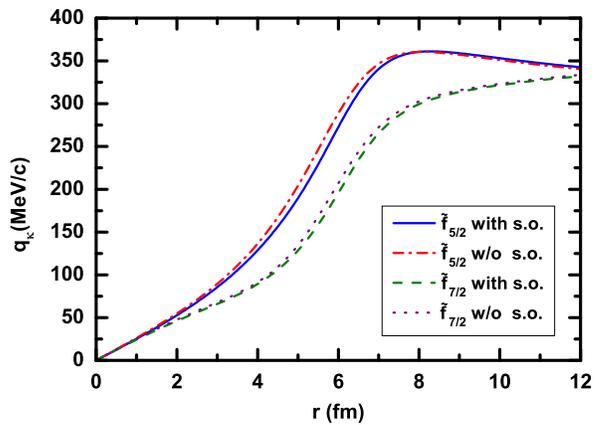}
\caption{(Color online)
Reduced superpotentials $q_\kappa(r)$ for the $\tilde{f}_{5/2}$ and $\tilde{f}_{7/2}$ blocks with and without spin-orbit (SO) term.
    \label{Fig1}}
\end{figure}

In order to obtain the SUSY partner Hamiltonian $\tilde{H}$ in Eq.~(\ref{Eq:Htil}), first of all, the differential equations (\ref{Eq:qr}) for the reduced superpotentials $q_\kappa(r)$ should be solved with the boundary condition $q_\kappa(0)=0$.
The solutions are shown in Fig.~\ref{Fig1} by taking the $\tilde{f}_{5/2}$ and $\tilde{f}_{7/2}$ blocks as examples.
In order to identify the SO effects, the corresponding results without SO term, i.e., letting $U(r)\equiv0$ from the beginning, are shown for comparison.
Since the SO potential mainly contributes around the nuclear surface, the asymptotic behaviors of $q_\kappa(r)$ at $r\rightarrow0$ and $r\rightarrow\infty$ are the same for a given $\kappa$ independent of the SO term.
In addition, it can be seen that the $q_\kappa(r)$ obtained with the SO term is slightly smaller than that obtained without SO term at the surface region. This is because the SO potential $U(r)$ is positive at this region.

\begin{figure}
\includegraphics[width=8cm]{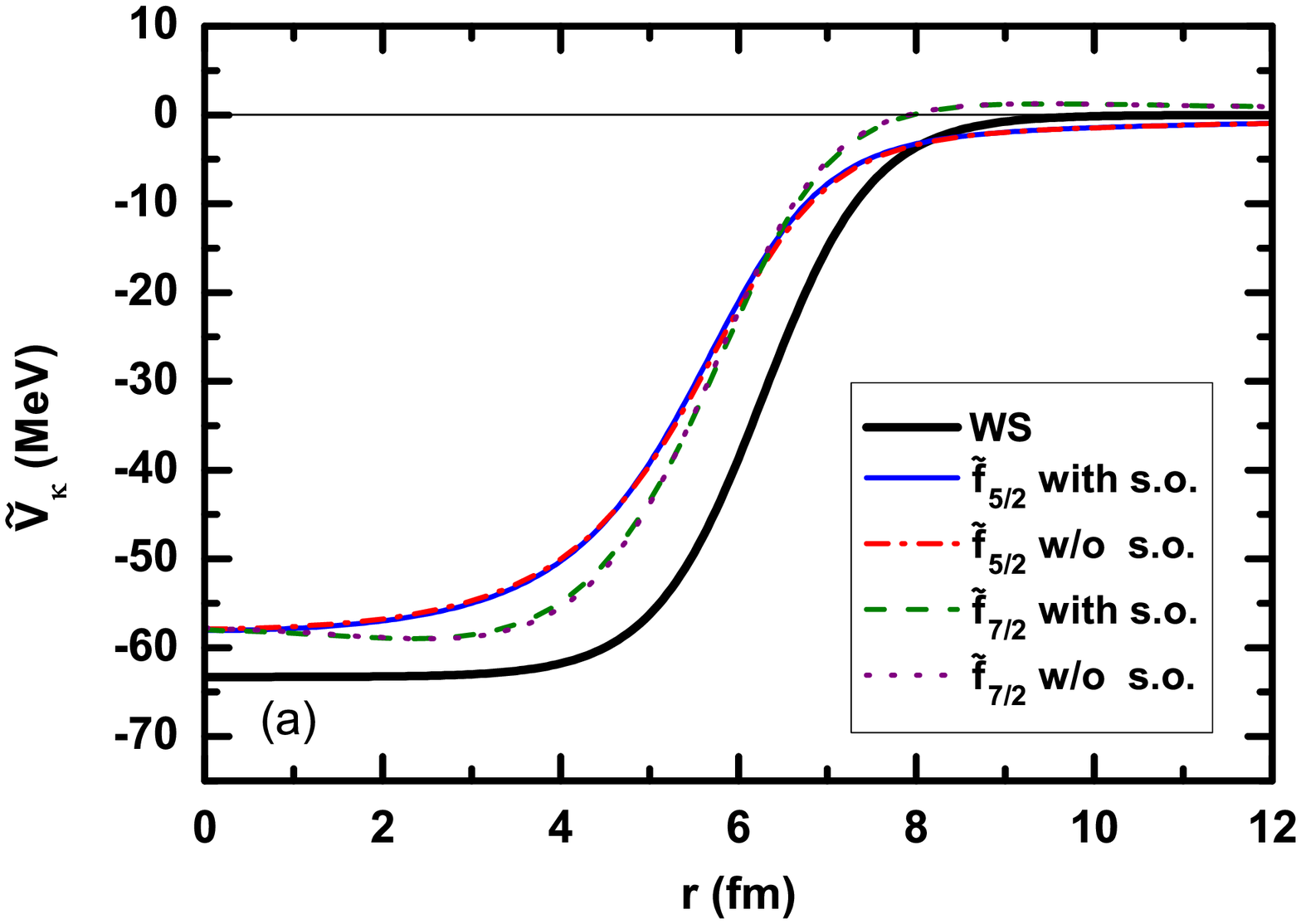}\\
\includegraphics[width=8cm]{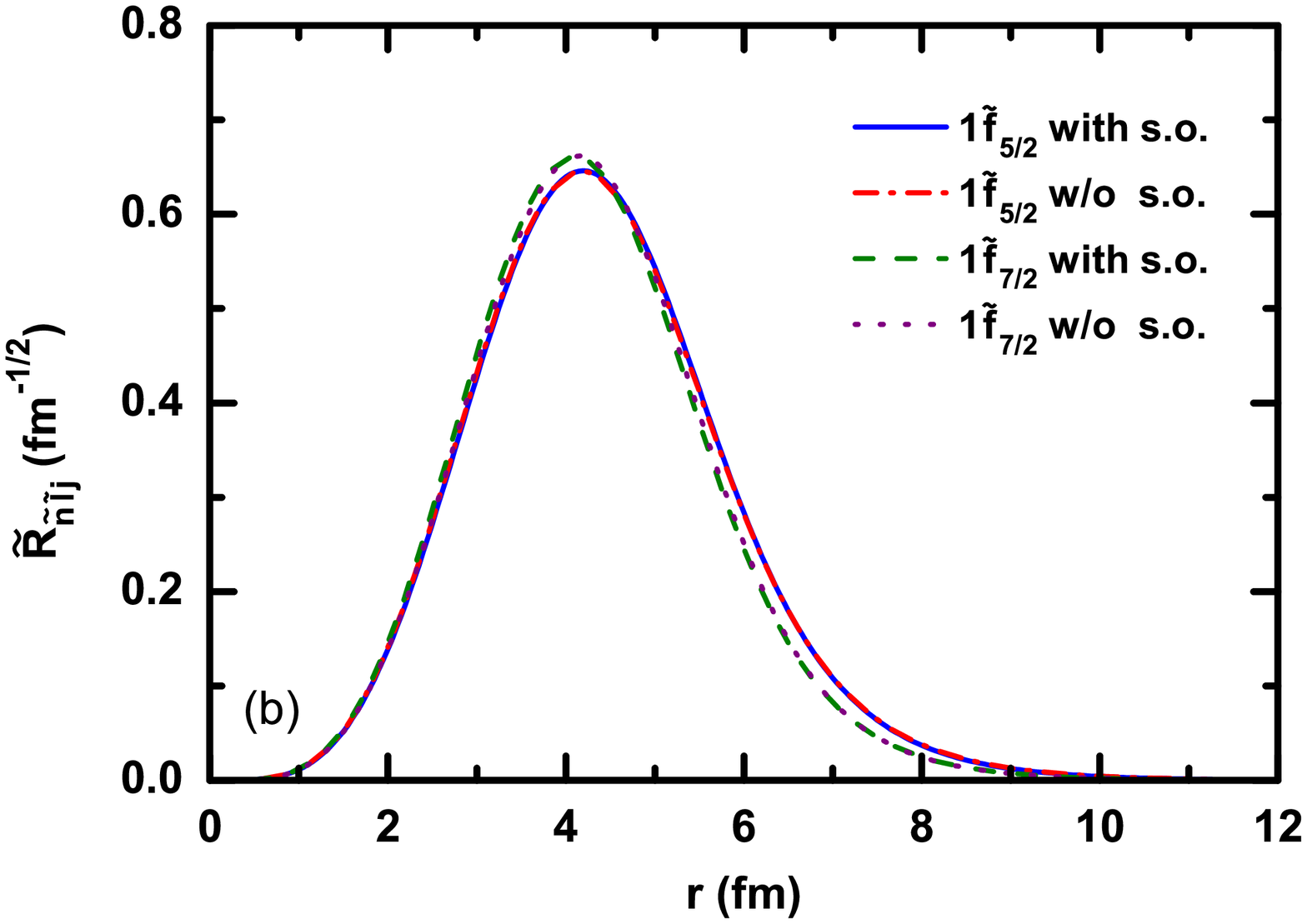}
\caption{(Color online) (a) Central potentials $\tilde{V}_\kappa(r)$ in $\tilde{H}$ for the $\tilde{f}$ block obtained with and without SO term. The Woods-Saxon potential in $H$ is shown with a thick line for comparison.
(b) The corresponding single-particle wave functions $\tilde{R}_{\tilde{n}\tilde{l}j}(r)$ of the $1\tilde{f}$ states.
    \label{Fig2}}
\end{figure}

In the upper panel of Fig.~\ref{Fig2}, the central potentials $\tilde{V}_\kappa(r)$ in $\tilde{H}$ for the $\tilde{f}_{5/2}$ and $\tilde{f}_{7/2}$ blocks are shown.
For comparison, the original Woods-Saxon potential $V(r)$ in $H$ is also shown as a thick line.
While the general features of $\tilde{V}_\kappa(r)$ have been discussed in Ref.~\cite{Liang2013}, here we focus on the effects due to the SO term.
Nevertheless, it is found that the SO effects on $\tilde{V}_\kappa(r)$ are almost invisible.
The same conclusion is also valid for the single-particle wave functions $\tilde{R}_{\tilde{n}\tilde{l}j}(r)$ thus obtained.
As shown in the lower panel of Fig.~\ref{Fig2}, the difference between the wave functions of pseudospin doublets ($1\tilde{f}_{5/2}, 1\tilde{f}_{7/2}$) is related to their $2.381$ MeV splitting in energy, while the wave functions obtained with and without SO term are almost identical.
Therefore, we will focus on the SO effects on the energy splitting of the pseudospin doublets.

\begin{figure}
\includegraphics[width=8cm]{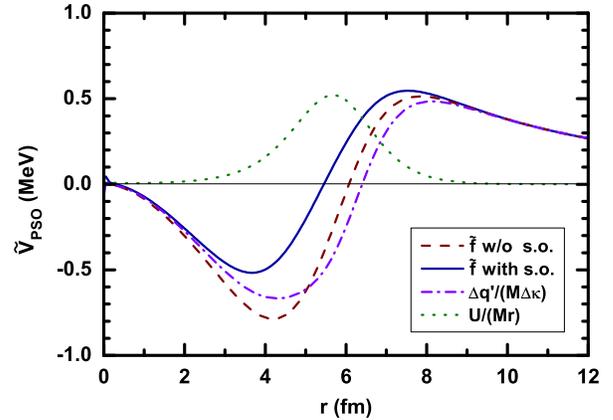}
\caption{(Color online) Pseudospin-orbit potentials $\tilde{V}_{\rm PSO}(r)$ for the $\tilde{f}$ block.
The symmetry breaking potential obtained with the SO term (solid line) is decomposed into the contributions from the first (dash-dotted line) and second (dotted line) terms on the right-hand side of Eq.~(\ref{Eq:VPSO}).
The symmetry breaking potential obtained without SO term is shown with short dotted line for comparison.
    \label{Fig3}}
\end{figure}

We first show the PSS-breaking potentials $\tilde{V}_{\rm PSO}(r)$ in Eq.~(\ref{Eq:VPSO}) in Fig.~\ref{Fig3} by taking the $\tilde{f}$ block as an example.
It is worthwhile to emphasize the main features of $\tilde{V}_{\rm PSO}(r)$~\cite{Liang2013}:
(1) these PSS-breaking potentials are regular functions of $r$;
(2) their amplitudes directly determine the sizes of reduced PSO splittings $\Delta E_{\rm PSO}\equiv(E_{j_<}-E_{j_>})/(2\tilde l+1)$;
(3) the shape of $\tilde{V}_{\rm PSO}(r)$ is negative at small radius but positive at large radius with a node at the surface region, which could explain the decrease of the PSO splittings with increasing single-particle energies, and even reverse as approaching the single-particle threshold.

In order to present the SO effects on the PSS-breaking potentials clearly, the $\tilde{V}_{\rm PSO}(r)$ obtained with the SO term are decomposed into the contributions from the first and second terms on the right-hand side of Eq.~(\ref{Eq:VPSO}), denoting as $\Delta q'/M\Delta\kappa$ and $U/Mr$, respectively.
These two terms can be regarded as indirect and direct effects of the SO term, respectively. The former one represents the SO effects on $\tilde{V}_{\rm PSO}(r)$ via the reduced superpotentials $q_\kappa(r)$ as shown in Fig.~\ref{Fig1}. The latter is nothing but the SO potential itself appearing in the original Hamiltonian $H$ in Eq.~(\ref{Eq:HkaNew}).

By comparison to the result obtained without SO term, the $\Delta q'/M\Delta\kappa$ term with SO term is raised for $r<5$~fm and lowered for $r>5$~fm. However, such effect is the order of $0.1\sim0.2$~MeV on $\tilde{V}_{\rm PSO}(r)$, and eventually less net influence on $\Delta E_{\rm PSO}$ due to the cancellation for $r<5$~fm and $r>5$~fm.
For the SO potential $U(r)/Mr$ shown with the dotted line in Fig.~\ref{Fig3}, appearing in the original Hamiltonian $H$ [Eq.~(\ref{Eq:HkaNew})] as well as the SUSY partner Hamiltonian $\tilde H$ [Eq.~(\ref{Eq:Htil})], it is always positive with a surface-peak shape.
It substantially raises the $\tilde{V}_{\rm PSO}(r)$, in particular for the surface region, and thus makes the reduced PSO splittings $\Delta E_{\rm PSO}$ systematically smaller.

\begin{figure}
\includegraphics[width=8cm]{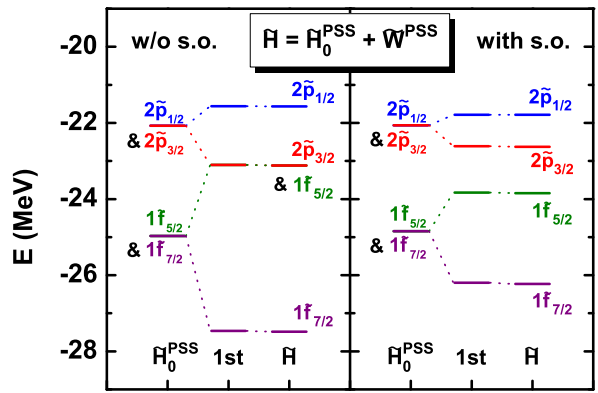}
\caption{(Color online) Single-particle energies for the pseudospin doublets $2\tilde p$ and $1\tilde f$ obtained at the exact PSS limit $\tilde H_0^{\rm PSS}$ and their counterparts obtained by the first-order perturbation calculations, as well as those obtained with the SUSY partner Hamiltonian $\tilde{H}$.
The cases without and with SO term are shown in the left and right panels, respectively.
    \label{Fig4}}
\end{figure}

The Rayleigh-Schr\"odinger perturbation calculations~\cite{Greiner1994} are then performed based on the PSS-conserving Hamiltonian $\tilde H_0^{\rm PSS}$ with the PSS-breaking perturbation $\tilde W^{\rm PSS}$.
In Fig.~\ref{Fig4}, we show the single-particle energies for the pseudospin doublets $2\tilde p$ and $1 \tilde f$ obtained at the exact PSS limit $\tilde H_0^{\rm PSS}$ and their counterparts obtained by the first-order perturbation calculations, as well as those obtained with the SUSY partner Hamiltonian $\tilde{H}$.
It is clear that the PSO splittings can be excellently reproduced by the first-order perturbation calculations for both cases without and with SO term.
This indicates the SO potential does not change the perturbative nature of PSS.

Quantitatively, the first-order perturbation corrections to the single-particle energies are simply expressed as $\kappa\int \tilde{R}^*_{\tilde{n}\tilde{l}j}(r)\tilde{V}_{\rm PSO}(r)\tilde{R}_{\tilde{n}\tilde{l}j}(r)dr$.
As shown in Fig.~\ref{Fig2}(b), the SO term has practically no effect on the single-particle wave functions $\tilde{R}_{\tilde{n}\tilde{l}j}(r)$.
Therefore, the SO effects on the PSO splittings mainly come from its effect on the PSS-breaking potentials $\tilde{V}_{\rm PSO}(r)$, where the so-called direct effect discussed above is found to be dramatic.
The PSO splittings of $2\tilde p$ and $1\tilde f$ doublets are reduced from $1.552$ and $4.364$~MeV to $0.840$ and $2.381$~MeV, respectively.

\begin{figure}
\includegraphics[width=8cm]{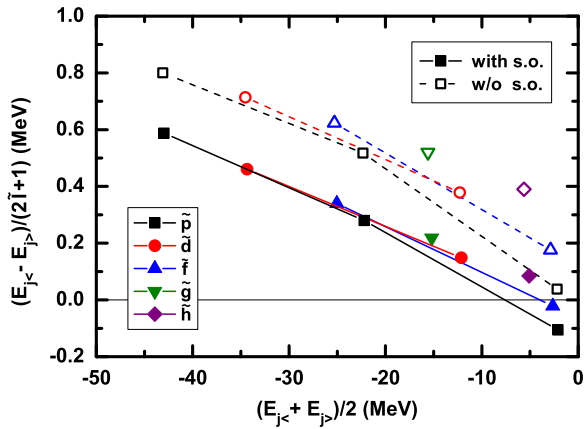}
\caption{(Color online) Reduced pseudospin-orbit splittings $(E_{j_<}-E_{j_>})/(2\tilde{l}+1)$ vs. the average single-particle energies $(E_{j_<}+E_{j_>})/2$.
The results obtained with and without SO term are shown with filled and open symbols, respectively.
    \label{Fig5}}
\end{figure}

Finally, as a general pattern, the reduced PSO splittings $\Delta E_{\rm PSO}$ for all bound pseudospin doublets are shown as a function of the average single-particle energies $E_{\rm av}=(E_{j_<}+E_{j_>})/2$ in Fig.~\ref{Fig5}.
The results obtained with and without SO term are shown with filled and open symbols, respectively.
The general tendency that the PSO splittings become smaller with increasing single-particle energies maintains. It is noted that after including SO term, the PSO splittings can even reverse as approaching the single-particle threshold. It is also seen that the slopes of $\Delta E_{\rm PSO}$ vs. $E_{\rm av}$ become slightly gentler with the SO term.
Furthermore, the SO term reduces the $\Delta E_{\rm PSO}$ systematically by $0.15\sim0.3$~MeV, and this effect can be understood in a fully quantitative way now.

%==================Summary and Perspectives==============================
\section{Summary and Perspectives}\label{Sect:IV}

The origin of pseudospin symmetry and its breaking mechanism are investigated by using the supersymmetric quantum mechanics and perturbation theory.
In the present work, we mainly focus on the effects caused by the spin-orbit term, which appears from the second-order correction in $1/M$ in SRG but plays a crucial role in nuclear shell structure.

The SO term shows both indirect and direct effects on the PSS-breaking potentials $\tilde{V}_{\rm PSO}(r)$.
The indirect effect due to the changes of the reduced superpotentials $q_\kappa(r)$ is rather small.
As a result, the central potentials $\tilde V_\kappa(r)$ in the SUSY partner Hamiltonian $\tilde H$ and the corresponding single-particle wave functions $\tilde{R}_{\tilde{n}\tilde{l}j}(r)$ obtained with and without SO term are almost identical.
In contrast, the direct effect corresponds to the SO potential itself appearing in both $H$ and $\tilde H$.
It reduces the PSO splittings $\Delta E_{\rm PSO}$ substantially.

The perturbation calculations are performed based on the PSS-conserving Hamiltonian $\tilde H_0^{\rm PSS}$ with the PSS-breaking perturbation $\tilde W^{\rm PSS}$.
The calculated results demonstrate the perturbative nature of PSS maintains even a substantial SO potential is included.
Finally, the SO effects on the PSO splittings are interpreted in a fully quantitative way.

One should take the intrinsic relations between the spin-orbit potential and the central potential or the effective mass into account to explain why PSS is conserved better than SS in realistic nuclei.
In this sense, PSS must be regarded as a relativistic symmetry, and it should be recognized in the Dirac equation, or equivalently the Schr\"odinger-like equation obtained by using the SRG technique.
For example, the so-called Darwin term in the SRG expansion, which is related to the effective mass of nucleon, should be investigated.
Works along this line are in progress.

\section*{ACKNOWLEDGMENTS}
This work was partly supported by the Major State 973 Program 2013CB834400,
the NSFC under Grants No. 11105005, No. 11105006, No. 11175002,
the China Postdoctoral Science Foundation Grant No. 2012M520101,
the Research Fund for the Doctoral Program of Higher Education under Grant No. 20110001110087,
and the Grant-in-Aid for JSPS Fellows under Grant No. 24-02201.

%\bibliography{PSS}

\end{CJK*}
\end{document}